\documentclass[10pt,aps,prx,twocolumn,amsmath,amssymb,floatfix,superscriptaddress]{revtex4-2}

\usepackage{caption}
\usepackage{graphicx}
\usepackage{nicefrac}
\usepackage{subcaption}
\usepackage{physics}
\usepackage[normalem]{ulem}
\usepackage{xcolor}
\usepackage[title]{appendix}
\usepackage{hyperref}
\usepackage{comment}

\newcommand{\affA}{Van der Waals-Zeeman Institute, Institute of Physics, University of Amsterdam, 1098 XH Amsterdam, Netherlands}
\newcommand{\affB}{QuSoft, Science Park 123, 1098 XG Amsterdam, the Netherlands}

\newcommand{\affD}{Institute for Quantum Electronics, ETH Z\"urich, Otto-Stern-Weg 1, 8093 Z\"urich, Switzerland}
\newcommand{\affE}{Quantum Center, ETH Z{\"u}rich, 8093 Z{\"u}rich, Switzerland}

\begin{document}

\preprint{APS/123-QED}

\title{Non-paraxial effects on laser-qubit operations}%

\author{L.~P.~H. Gallagher}\affiliation{\affA}
\author{M. Mazzanti}\affiliation{\affD}\affiliation{\affE}
\author{Z.~E.~D.~Ackerman}\affiliation{\affA}
\author{A. Safavi-Naini}\affiliation{\affA}\affiliation{\affB}
\author{R. Gerritsma}\affiliation{\affA}\affiliation{\affB}
\author{R.~J.~C. Spreeuw}\affiliation{\affA}\affiliation{\affB}

\date{\today}

\begin{abstract}
Tightly-focused laser beams, or optical tweezers, are essential for analogue and digital quantum simulation with neutral atoms and trapped ions. Despite this, most of the current intuition and theoretical treatment utilizes the paraxial approximation, which breaks down at the focus of optical tweezers. We develop an analytic model, which we use in tandem with numerical simulations, to quantify how non-paraxial effects will manifest in the next-generation of scalable quantum hardware, where tightly focused beams are used for individual qubit control. In particular, we calculate the light potentials of Gaussian and Laguerre-Gaussian beams driving the quadrupole $^2$S$_{1/2}\rightarrow$ $^2$D$_{5/2}$ transition in $^{40}$Ca$^+$. Longitudinal field components in the beam center cause spatially-dependent Rabi frequencies and AC Stark shifts, leading to unexpected qubit-motion coupling. 
We characterize single- and two-qubit gate infidelities due to this effect with an analytic model and numerical simulation. We identify regimes where non-paraxial effects should be taken into account for high-precision quantum control. Finally, we highlight that non-paraxial effects are potentially more severe in the case of neutral atom and molecule addressing.

\end{abstract}

\maketitle

\section{Introduction}

The application of optical tweezers in cold atomic physics has made remarkable progress in the past decade. On the one hand, optical tweezers can hold thousands of neutral atoms to form large arrays of atomic qubits, paving the way for the construction of a scalable quantum computer~\cite{Barredo:2016, Endres:2016, Omran:2019, Graham:2019, Kaufman:2021,Urech:2022,bluvstein:2024,Pause:24,Manetsch:2024}. On the other hand, in trapped-ion platforms for quantum simulation and computation, tightly focused laser beams are used to address individual ions and prepare, manipulate and measure their internal (qubit) state~\cite{Schindler:2013,Wright:2019,peshkov:2023,Lange:2022}. Furthermore, optical tweezers may be used to modify the local confinement of trapped ions~\cite{Olsacher:2020,Teoh:2021,Espinoza:2021,Mazzanti:2021,Stopp_ang_mom2022,Schwerdt:2024,vasquez:2024} with applications in quantum computing and simulation. As the ions are typically separated by several $\mu$m~\cite{James:1998}, it is crucial to obtain the smallest possible foci for the addressing lasers to avoid cross talk \cite{Parrado:2021}.

\begin{figure*}[t]
    \centering
    \includegraphics[width=\linewidth]{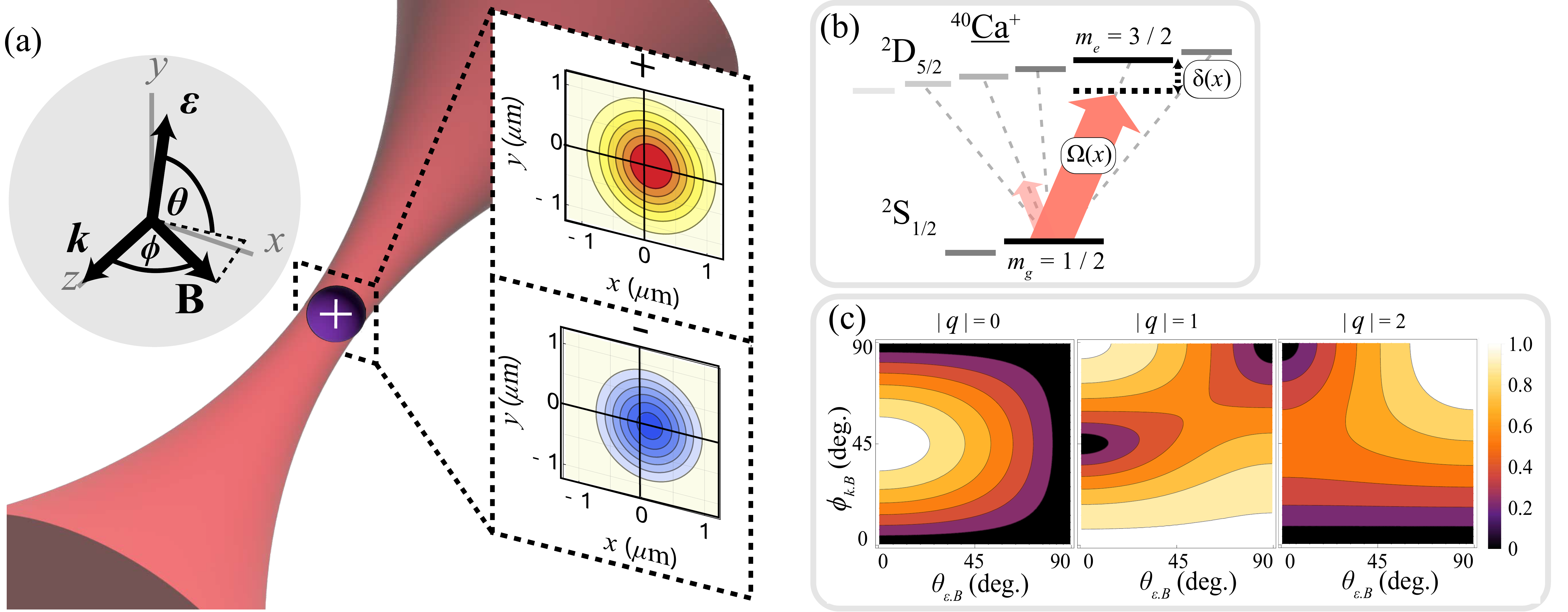}
    \caption{\textbf{Tweezer setup and orientation.} (a) We consider a tightly-focused beam resonantly driving a quadrupole transition in a trapped ion. In the grey circle inset we illustrate the coordinate system, with angle $\phi$ describing the angle between the magnetic field $\bold{B}$ and propagation vector $\hat{k}$, and angle $\theta$ the orientation of the polarization $\hat{\varepsilon}$ vector with respect to the projection of $\bold{B}$ on the transverse plane. For the case in which $\bold{B}\parallel\hat{\varepsilon}\parallel y$ (as studied below), we show the transverse potentials of the dressed qubit states \(\{\ket{+},\ket{-}\}\), which are displaced in the $x$ direction by approximately $\lambda/2\pi$. An off-axis field gradient and curvature at the ion's position can induce a qubit-motion coupling error. (b) Simplified level scheme of $^{40}$Ca$^+$ showing the resonant quadrupole transition with a spatially-varying Rabi frequency $\Omega(x)$. Off-resonant dipole and quadrupole coupling causes a spatially-varying AC Stark shift $\delta(x)$. (c) Geometric dependence of the quadrupole coupling ($m_g\rightarrow m_g+q$ for $q=0,\pm1,\pm2$) at the beam center, as a function of angles \(\theta\) and \(\phi\), with the beam waist $w_0=\lambda=729$~nm. Light (dark) shadings signify high (low) coupling strength.}
    \label{fig1}
\end{figure*}

In the above applications, the tweezer light fields are typically considered to be Gaussian and described within the paraxial approximation. In this approximation, the divergence angle of the light field with respect to the optical axis is considered to be small.
However, close to the focus, large deviations from the paraxial approximation appear~\cite{Wang:2020,Spreeuw:2020,Spreeuw:2022,Unnikrishnan:2024, Thompson:2013}. These may cause strong gradients in the polarization components along the tweezer propagation direction, which in turn leads to the appearance of state-dependent forces in the direction perpendicular to the tweezer~\cite{Spreeuw:2020,Mazzanti:2023,Cui:2025,Mai:2025}.

In quantum simulation platforms based on atomic qubits these effects may lead to a loss of qubit coherence caused by qubit-motion entanglement~\cite{Leibfried:1996,Christandl:2016,Sutherland:2022}. In particular, in trapped ion crystals, these optical forces typically point along the weakly confined axial direction. Since the axial modes are harder to cool due to their broad spectral range~\cite{James:1998}, non-paraxial effects must be taken into account carefully. To this end, we present the first detailed study of the effects of the breakdown of the paraxial approximation in the case of an optical qubit encoded in a $^{40}$Ca$^+$ ion. 

In the following, we calculate the coupling of the tweezer both on the $^2$S$_{1/2}\rightarrow {}^2$D$_{5/2}$ quadrupole transition and the strongest dipole transitions in $^{40}$Ca$^+$. This enables us to determine spatially-dependent Stark shifts on the qubit states~\cite{Haffner:2003a}. We calculate the fidelity loss for single qubit gates on a single trapped ion due to the tweezer-induced qubit-motion coupling. Moreover, we consider the errors incurred while performing two-qubit operations, including an additional qubit-qubit interaction. We show that while a small loss of gate quality is expected, the fidelity should remain within the bounds of fault tolerance assuming typical experimental parameters~\cite{Bermudez:2017}. Our calculations may be straightforwardly extended to ions with similar level schemes such as Sr$^+$ and Ba$^+$, and the techniques presented may also be applied to neutral atoms and molecules.

\section{Results}

We consider a trapped $^{40}$Ca$^+$ ion addressed by a tightly-focused beam which resonantly drives the quadrupole transition between the qubit states $\ket{0}=4^2$S$_{1/2}$ ($m_g=1/2$) and $\ket{1}=3^2$D$_{5/2}$ ($m_e=3/2$), as shown in Fig. \ref{fig1}. The system is governed by the Hamiltonian 
\begin{equation}
    H=H_{\text{AL}} + H_\text{Z} + H_{\alpha}, 
    \label{hamiltonian1}
\end{equation}
where $H_{\text{AL}}$ is the atom-light Hamiltonian that includes the dipole and quadrupole interactions and $H_{\alpha}$ includes the off-resonant dipole coupling terms. Here $H_{\rm Z}
=g_j\mu_B\bold{B}\cdot\bold{\hat{J}}$ is the Zeeman Hamiltonian with \(g_j\) the Lande g-factor, \(\mu_B\) the Bohr magneton, and \(\bold{B}\) the magnetic field which sets the quantization axis. Finally, \(\bold{\Hat{J}}\) is the vector of angular-momentum operators for a particular angular-momentum manifold. 

We must go beyond the paraxial approximation to fully account for the effect of tightly focused beams on the fidelity of single and two-qubit gates. In the paraxial approximation, the electric field of a Gaussian beam is transverse to the direction of propagation and uniformly polarized. However, in the tightly-focused regime, longitudinal field components appear and the polarization becomes strongly position-dependent. In Eq.~\eqref{eq:E-1st-order} we provide an analytic expression for the electric field containing first order corrections to the paraxial approximation. We note that in an extremely tightly-focused regime one should instead use the exact electric field patterns which can be obtained numerically \cite{Novotnoy:2012,Spreeuw:2020}.

\begin{figure*}[t]
    \centering
    \includegraphics[width=\linewidth]{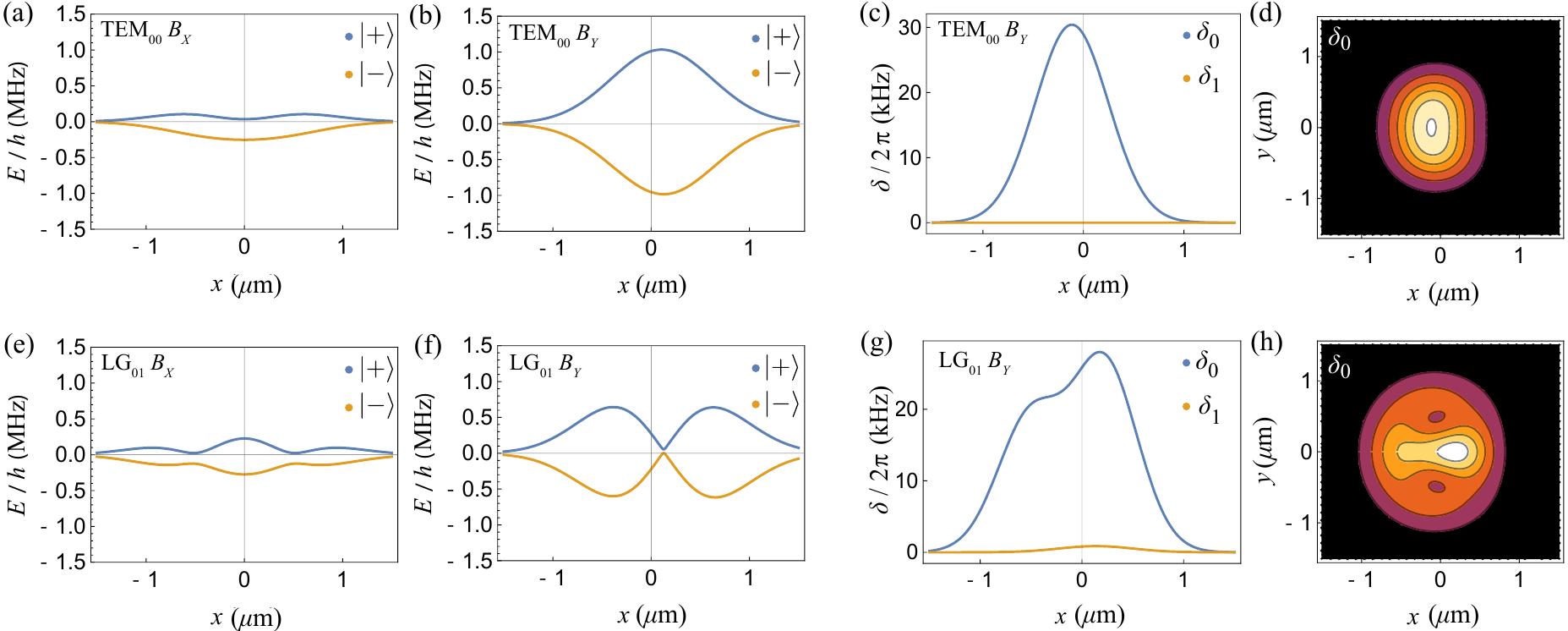}
    \caption{\textbf{Potential and light shifts.} Potential of the dressed qubit states $\{\ket{+},\ket{-}\}$ at the focus of an optical tweezer resonantly driving the quadrupole \(^2\text{S}_{1/2}\rightarrow\) \(^2\text{D}_{5/2}\) transition in $^{40}$Ca$^+$. (a) \& (b) Potential for a Gaussian beam $\text{TEM}_{00}$ as a function of \(x\) (with $y=0$ and $z=0$), with linear polarization in the \(y\) direction, and a magnetic field $|\bold{B}|=5$~G in the \(x\) and \(y\) directions. The laser parameters are $P_0=10$~$\mu$W and $w_0=729$~nm. (c) AC Stark shifts of the qubit states as a function of \(x\), with magnetic field $B_y$. (d) The dominant Stark shift as a transverse ($x-y$) contour plot in the focal plane ($z=0$). Light (dark) regions signify a large (small) Stark shift. (e) - (h) as above for a Laguerre-Gaussian beam $\text{LG}_{01}$.}
    \label{potential plots}
\end{figure*}

In the following, we show that the resulting spatially-dependent Rabi frequencies and AC Stark shifts between the qubit states, denoted $\Omega(x)$ and $\delta(x)$ respectively, generate residual qubit-motion coupling which adversely affects the fidelity of single- and two-qubit gates. The errors due to the curvature of the Rabi frequency have been previously studied~\cite{Sutherland:2022, West:2021}. Here, we quantify errors due to spatial variations in the Rabi frequency and Stark shift which uniquely arise in the non-paraxial regime.

The spatial form of $\Omega(x)$ and $\delta(x)$ depend on the optical light potentials $E(x)$. In Fig.~\ref{potential plots} we illustrate these potentials in the dressed basis $\{\ket{+},\ket{-}\}$, where $\ket{+}=c_1\ket{0}+c_2\ket{1}$ and $\ket{-}=c_1\ket{1}-c_2\ket{0}$ are linear superpositions of the qubit states (see methods). The potential is primarily due to the Rabi frequency between $\ket0$ and $\ket 1$, which means that $E(x)$ and $\Omega(x)$ have the same spatial profile. For resonant $\mathbf B= B_y \hat y$, the resulting profile is Gaussian but is shifted from the center of mass of the atom. In the regime where $w_0/\lambda\gtrsim 1$ the Rabi frequency maximum is displaced by an amount $x_0=\lambda/2\pi$ (see Fig.~\ref{magnus effect displacements}).  We repeat the same procedure for a Laguerre-Gauss mode \(\text{LG}_{p,l}=\text{LG}_{01}\) with $B= B_y \hat y$ and observe a similar shifted potential (see Fig. \ref{potential plots}(e)-(h)). We attribute this displacement to the circularly-polarized light field components which arise from the strong curvature of the longitudinal field near the focus of the tweezer in the non-paraxial regime. This is a generalization of the optical Magnus effect for a dipole transition in Ref.~\cite{Spreeuw:2020} to the case of a quadrupole transition. Finally we plot the AC Stark shifts (derived in the methods) \(\delta_i(x)\)  in \ref{potential plots}(d). We note that the spatial profile is different from the Rabi frequency and in particular, the displacement $x_0$ is in the opposite direction.

For an ion placed at $x_0$, with $\Omega^{(1)}(x_0) \equiv \frac{d\Omega(x)}{dx}\vert _{x=x_0}=0$, the system is described by the Hamiltonian 
\begin{align} 
H_3=&\frac{\hat{p}^2}{2m}+\frac{1}{2}m\omega^2\hat{x}^2+\hbar\left(\delta^{(0)}+\delta^{(1)} \hat{x}+\delta^{(2)}\hat{x}^2\right)\hat{\sigma}_z\notag\\
&+\hbar\left(\Omega^{(0)}+\Omega^{(2)}\hat{x}^2\right)\hat{\sigma}_x
\label{hamiltonian3}
\end{align}
where we have expanded $\delta(x)$ and $\Omega(x)$ about $x_0$, 
\(\delta(\hat{x})=\delta^{(0)}+\delta^{(1)} \hat{x}+\delta^{(2)}\hat{x}^2\) and \(\Omega(\hat{x})=\Omega^{(0)}+\Omega^{(2)}\hat{x}^2\) and have ignored a global Stark shift by imposing \(\delta_0(\hat{x})=-\delta_1(\hat{x})\). Here, \(\hat{x}=l_{\text{ho}}(\hat{a}+\hat{a}^{\dagger})\) and \(\hat{p}=i(\hbar/2l_{\text{ho}})(\hat{a}^{\dagger}-\hat{a})\), $\hat a\,(\hat a^\dagger)$ is the annihilation (creation) operator with \(l_{\text{ho}}=\sqrt{\hbar/2m\omega}\) the characteristic length of the harmonic oscillator.

\subsection{Single-qubit errors}

\begin{figure*}[t]
    \centering
    \includegraphics[width=0.85\linewidth]{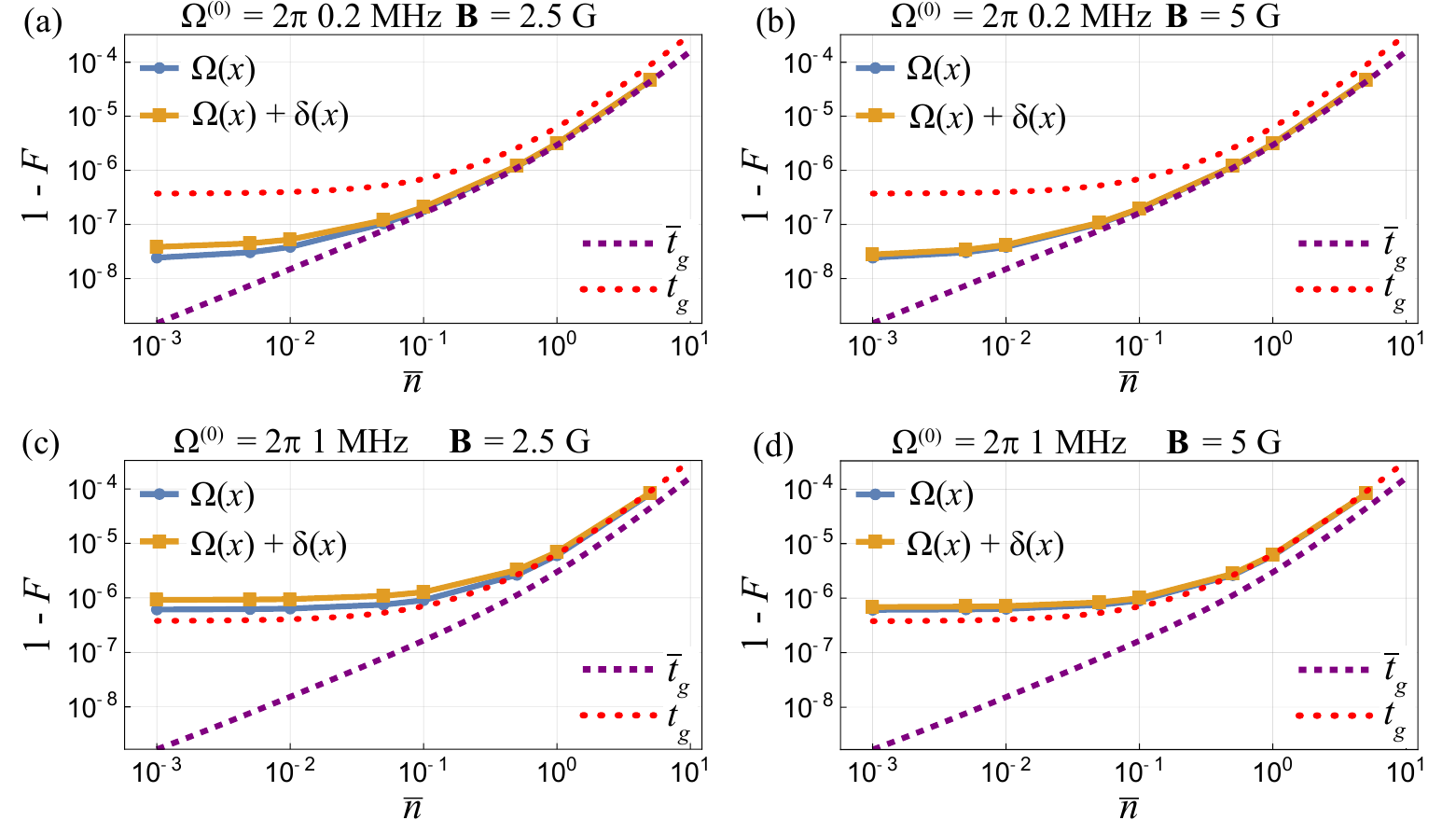}
    \caption{\textbf{Single-qubit errors.} Simulated infidelity $1-F$ for a single-qubit $\pi/2$ gate as a function of average occupation \(\Bar{n}\). We compare simulated errors due to the non-zero components of the Rabi frequency $\Omega(x)$ (blue points) and additionally Stark shift $\delta(x)$ (orange points), with the ion placed at the peak of the Rabi frequency $x_0$ and using a gate time $\Bar{t}_g$. The dashed lines are analytic expressions for the infidelity with a compensated (purple dashed) and uncompensated (red dotted) gate time. We consider Rabi frequency amplitudes $\Omega^{(0)}=2\pi~0.2$~MHz and $2\pi~1$~MHz, and magnetic field strengths $|\textbf{B}|=2.5$~G and $5$~G. Note the range of temperatures lies below typical Doppler-cooling limits.
    Decreasing the magnetic field linearly increases the error due to the Stark shift $\delta(x)$, making it more prominent. Increasing the Rabi frequency amplitude increases off-resonant errors which are excluded in the analytic model, which dominate at low $\bar{n}$.}
    \label{fig:fidelity}
\end{figure*}

We apply two unitary transformations on $H_3$ (see the Supplemental Material) to eliminate the linear and quadratic qubit-motion coupling induced by the Stark shift components $\delta^{(1)}$ and $\delta^{(2)}$ respectively. Here we neglect off-resonant motional processes that change the Fock state. The transformed Hamiltonian $\tilde{H}$ in angular frequency units is,
\begin{equation}
\tilde{H}=H_{\text{id}}+\kappa_z(\hat{n}+1/2)\hat{\sigma}_z+\kappa_x(2\hat{n}+1)\hat{\sigma}_x,
\label{eq:ham1qubit}
\end{equation}
where \(H_{\text{id}}=\omega(\hat{a}^{\dagger}\hat a+1/2)+\Omega^{(0)}\hat{\sigma}_x\) is an ideal \(\hat{\sigma}_x\) qubit rotation.
The second term can be interpreted as the error from squeezing the ion's harmonic motion due to the Stark shift curvature, labeled with error parameter \(\kappa_z=2\delta^{(2)}l_{\text{ho}}^2\). The third term describes \(\sigma_x\)-type errors due to the Rabi frequency curvature and the Stark shift gradient, labeled \(\kappa_x=2\Omega^{(0)}(\delta^{(1)})^2l_{\text{ho}}^2/\omega^2+\Omega^{(2)}l_{\text{ho}}^2\). The physical contributions of the error parameters is discussed in detail in the methods section. The fidelity of the resulting single-qubit gate is given by \cite{Nielsenprocessfidelity}, 
\begin{align}
\Bar{F}=\frac{\sum_l \text{tr}[\hat{U}_{\text{id}}\hat{\sigma}^{\dagger}_l\hat{U}_{\text{id}}^{\dagger}\boldsymbol{\hat{\sigma}}_l(\hat{U}_{\text{real}})+d^2]}{d^2(d+1)}, 
\label{fidelity}
\end{align}
where $\boldsymbol{\hat{\sigma}}_l(\hat{U}_{\text{real}})\equiv \text{tr}_{\text{Fs}}\left(\hat{U}_{\text{real}}(\ket{n}\bra{n}\otimes\hat{\sigma}_l)\hat{U}_{\text{real}}^{\dagger}\right)$ is the projector on Fock state $\ket{n}$, and $d=2$ \cite{Mazzanti:2021}. The unitaries are \(\hat{U}_{\text{id}}=\exp(-iH_{\text{id}}t)\) and \(\hat{U}_{\text{real}}=\exp(-i\tilde{H}t)\), and the motional modes are weighted with the Bose-Einstein distribution $P_{\text{th}}(n,\Bar{n})=\Bar{n}^n/(1+\Bar{n})^{n+1}$, where the average occupation number \(\Bar{n}=1/(e^{\frac{\omega\hbar}{k_B T}}-1)\) is determined by the temperature of the ion crystal, \(T\). 

We expand \(\Bar{F}\) to second order in \(\kappa_x\) and \(\kappa_z\), and neglect cross terms, to arrive at an analytic expression for the infidelity,

\begin{equation}
    1-\Bar{F}\approx \frac{1}{6(\Omega^{(0)})^2}(1+8\Bar{n}(\Bar{n}+1))\left(\pi^2\kappa_x^2+\kappa_z^2\right),
    \label{analytic fid}
\end{equation}
when using a gate time \(t_g=\pi/2\Omega^{(0)}\) (the ideal gate time for a $\pi/2$ pulse with $H_{\text{id}}$).

The error due to non-zero $\kappa_x$ is at its core an error in the timing of the qubit rotation and can be partially compensated. To this end, we define a new gate time \(\bar t_g=\pi/2\Bar{\Omega}\) where $\bar \Omega$ is the thermally-averaged Rabi frequency (known for a given $\bar{n}$), and provide an expression for the compensated gate fidelity in the methods section (Eq. \eqref{analytic fid appendix 2}). This expression allows us to significantly reduce the infidelity without numerical optimization of the gate time. The fidelity calculated with both gate times is compared against numerical simulation in Fig. \ref{fig:fidelity}. Furthermore, the dependence of the fidelity on the beam waist is considered in the Supplemental Material.

\subsubsection{Numerical simulation}

We compute the fidelity numerically with the Hamiltonian in Eq.~(\ref{hamiltonian3}). In Figure~\ref{fig:fidelity}, we show the infidelity evaluated for the parameters in the previous section, with an axial trap frequency \(\omega_{\text{ax}}=2\pi~0.5\)~MHz corresponding to  \(l_{\text{ho}}=16\)~nm, with two Rabi frequency amplitudes \(\Omega^{(0)}=2\pi~0.2\)~MHz and $2\pi~1$~MHz, and two magnetic field strengths $|\bold{B}|=2.5$~G and $5$~G. For the case \(\Omega^{(0)}=2\pi~1\)~MHz and $|\bold{B}|=5$~G, the field gradients and curvatures at the position of the ion (at $x_0\sim 108$~nm from the beam center) are \(\delta^{(1)}\approx 2\pi~20\)~Hz/nm, \(\delta^{(2)}\approx 2\pi~0.03\)~Hz/nm$^2$ and \(\Omega^{(2)}\approx 2\pi~2\)~Hz/nm$^2$, corresponding to fractional error parameters \(\kappa_x/\Omega^{(0)}\approx5\times10^{-4}\) and \(\kappa_z/\Omega^{(0)}\approx1\times10^{-5}\).  
We find that $\Omega^{(2)}$ is the dominant error source, as evidenced by the small difference between the solid orange and blue lines, which include and neglect $\delta(x)$, respectively. In the methods we describe parameter regimes where the non-paraxial error due to $\delta^{(1)}$ dominates.

Finally, we compare the analytic expression in  Eq. \eqref{analytic fid appendix 2} with gate time of $\bar t_g$ to the numerical results. We attribute the discrepancy between the analytic and the numerical results at larger $ \Omega^{(0)}$ to off-resonant processes which we neglected in our analytic treatment. Finally, we note that the presence of the zero-point energy means that even with the exclusion of off-resonant processes the infidelity does not tend to $0$ as $\Bar{n}\rightarrow0$. In this regime, the performance of the gate is significantly improved by correcting the gate timing as described above (compare the dashed and dotted lines in Fig.~\ref{fig:fidelity})

\subsection{Two-qubit errors}

We further extend our description to addressing two ions in a $N$-ion chain, in order to perform parallel single-qubit operations or two-qubit gates. The system is described by the eigenmodes of the collective motion of the chain, with the position operator for ion $k$: $\hat{x}_k=\sum_m l_mb_{mk}(\hat{a}_m+\hat{a}^{\dagger}_m)$ with characteristic length $l_m=\sqrt{\hbar/2m\omega_m}$ for normal mode $m$, and eigenvector $b_{mk}$ for the $k$th ion with the $m$th mode. 

We describe addressing two ions simultaneously in order to perform single-qubit gates in the methods section. We calculate qubit-motion coupling on all motional modes, alongside an additional qubit-qubit interaction term. For the parameter set used previously the qubit-qubit interaction is on the Hertz level, and thus negligible compared to the individual qubit-motion coupling errors.

\subsubsection{M{\o}lmer-S{\o}rensen gate}

Two-qubit gates can be implemented by amplitude-modulating the tweezers with intensity $A(t)=\frac{1}{2}(1-\cos(\nu t))$, with $\nu=\omega_r+\mu$. Here the gate uses the radial motion of the ions, with the detuning $\mu$ given with respect to the radial trap frequency $\omega_r$. The gate implements interactions $g_{r}\hat{\sigma}_x^{(i)}\hat{\sigma}_x^{(j)}$ between ions $i$ and $j$ with $g_{r}\sim \eta_{mi}\eta_{mj}(\Omega^{(0)})^2/\mu$ where $\eta_{mj}=b_{mj}\eta=b_{mj}k\sqrt{\hbar/2m\omega_r}$~\cite{Molmer:1999,Sackett:2000,Roos:2007,Benhelm:2008b,Kim:2009}.

\begin{figure}[t]
    \centering
    \includegraphics[width=\linewidth]{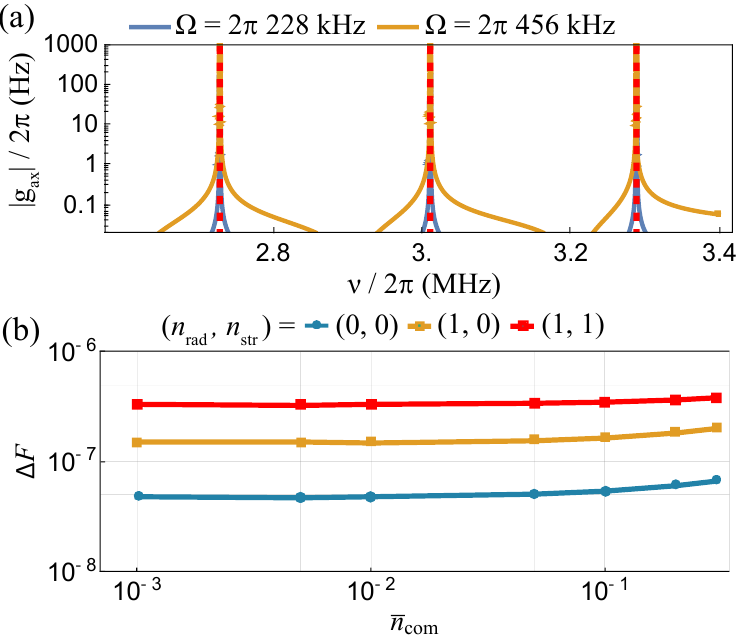}
    \caption{\textbf{Two-qubit errors.} (a) Magnitude of the axial qubit-qubit interaction $|g_{ax}|$ induced during the M{\o}lmer S{\o}rensen gate as a function of laser modulation frequency $\nu$. We consider the axial modes (red dashed lines) for a 10-ion chain with trap frequencies $(\omega_r,\omega_z)=2\pi(3.0,0.5)$~MHz. The interaction strength is shown for the Rabi frequency used in simulations $\Omega=2\pi~228$~kHz (blue) and twice this value $\Omega=2\pi~456$~kHz (orange). (b) The fidelity difference $\Delta F$ due to the non-zero tweezer gradient $\delta(x)$ and curvature $\Omega(x)$ while performing a M{\o}lmer-S{\o}rensen gate, as a function of average occupation in the axial centre-of-mass mode $\bar{n}_{\text{com}}$. We consider different Fock state occupations in the radial and axial stretch modes. The gate parameters are $\Omega^{(0)}=2\pi~228$~kHz, $\mu=2\pi~50$~kHz and $\eta=0.056$, and the magnetic field $\bold{B}=2.5$~G lies in the $y$ direction of the tweezers. Here we consider a Fock state cut-off $(n_{\text{com}},n_{\text{str}},n_{\text{rad}})=(12,8,12)$ limited by computational cost.}
    \label{two_qubit_figure}
\end{figure}

We note that resonant enhancement may occur if $\nu$ is close to one of the axial mode frequencies. In this case, qubit-qubit interactions of order $g_{\text{ax}}\sim(\delta^{(1)})^2l_m^2/\mu_m$ occur, with $\mu_m$ the detuning from the $m$th axial mode. For the parameters considered here, this amounts to $\lesssim 2\pi$~1~kHz$^2/\mu$. For long ion crystals with numerous axial modes coincidence of $\nu$ and $\omega_m$ is more likely. To illustrate we show the magnitude of the induced interaction $|g_{\text{ax}}|$ as a function of laser modulation frequency $\nu$ in Fig. \ref{two_qubit_figure}(a). Note that the interaction strength increases drastically with the Rabi frequency, as $g_{\text{ax}}\propto\Omega^4$.
We conclude that careful choice of parameters will keep the unwanted qubit-qubit interaction to the sub-Hertz level.

We perform a numerical simulation of the M{\o}lmer-S{\o}rensen gate for a two-ion chain. We determine the gate fidelity using Eq. \eqref{fidelity} both including (denoted $F_{\text{real}}$) and excluding (denoted $F_{\text{ideal}}$) the tweezer gradient and curvature. In Fig. \ref{two_qubit_figure}(b) we plot the fidelity difference $\Delta F=F_{\text{ideal}}-F_{\text{real}}$ (the error arising purely from the beam gradient and curvature) against the average thermal occupation $\bar{n}_{\text{com}}$. We find that the M{\o}lmer-S{\o}rensen gate is relatively robust to the considered errors, even with thermal occupation in the radial modes. We conclude that the errors considered here are significantly smaller than typical experimental errors due to laser stabilization or beam alignment, and thus do not limit the performance of two-qubit gates on current setups.

\section{Discussion}
\label{discussion}

In this paper we have considered the non-paraxial effects of light when addressing trapped ions with optical tweezers. The spatial profile of the potential of Gaussian and Laguerre-Gaussian beams are positioned off-axis with respect to the beam intensity. As a result, when the ion is placed at the peak of the Rabi frequency $\Omega(x)$, it experiences a non-zero curvature $\Omega(x)$, and a non-zero gradient of the Stark shift $\delta(x)$. We evaluated the size of the errors resulting from the non-zero curvature and gradient terms when performing single- and two-qubit gates, which manifest as qubit-motion entanglement. Our numerical simulations, as well as analytic results, show that while the size of the error is strongly dependent on the trapping frequency and the ion temperature, non-paraxial errors are generally very small when using typical experimental parameters. Moreover, suitable choice of laser polarization and magnetic field orientation can surpress the errors considered here.

We highlight that non-paraxial effects are potentially more severe when addressing neutral atoms and molecules. The resulting qubit-motion coupling has been observed as a source of qubit decoherence in Refs. \cite{Unnikrishnan:2024,Thompson:2013,Ammenwerth:2024}. For the case of a resonant tweezer our analysis may be straightforwardly applied to the neutral atom case. Here the linear qubit-motion coupling parameter $\zeta=\delta^{(1)}l_{\text{ho}}/\omega$ would be larger due to the weaker confinement $\omega$ compared to ions in a Paul trap. In this case, the single-qubit gate infidelity calculated from Eq. \ref{analytic fid} could increase drastically, as $(1-\Bar{F})\propto\zeta^4$. Note that the coupling also depends on the tweezer parameters, the atomic transition, and the orientation of the magnetic field and polarization. For the case of off-resonant tweezers - for instance used to confine neutral atoms, non-paraxial effects could also arise. This would depend on the atomic polarizability in the tweezer, and is not considered in this work. Furthermore, other error sources could arise at the scale of the non-paraxial effects considered here. An example of this is intensity fluctuations in the optical tweezer on the spatial scale of the ion's wavepacket \cite{Cetina:2022,Wineland:1998}. 

In conclusion, the errors considered here lie below the typical thresholds for quantum error correction \cite{Raussendorf:2007}. Compared to errors due to beam misalignment or laser stabilization, non-paraxial effects are not a limiting factor on current experimental setups. However, at low temperatures and high laser intensities non-paraxial effects can be a distinguishable error source. The effects can be especially difficult to compensate, due to being an inherent feature of the light field. This may be important for the next-generation of ion-trap quantum computers, which strive towards lower temperatures and utilize individually-addressing, high-intensity beams \cite{Pogorelov:2021,Schwerdt:2025,Zhao:2025}. Furthermore, our theoretical treatment can be applied towards future work, for instance a detailed study of non-paraxial effects on neutral atoms and molecules \cite{Blodgett:2025}, and for engineering quantum logic gates which use non-paraxial effects, as recently experimentally realized \cite{Cui:2025,Mai:2025}.

\section{Methods}

\subsection{Atom-light interactions in the non-paraxial regime}

A simple way to obtain a first-order correction to the paraxial approximation \cite{Davis:1979,verde2023trapped} is to start from the vector potential $\mathbf{A}(\mathbf{r},t)$ which satisfies the Lorenz condition $\nabla\cdot\mathbf{A}+(1/c^{2})\partial\Phi/\partial t=0$. The corresponding $\mathbf{E}(\mathbf{r},t)$ can be expressed entirely in terms of $\mathbf{A}(\mathbf{r},t)$. Neglecting the spatial derivatives of higher than first order, the electric field in the focal plane $z=0$ is given by\cite{Novotnoy:2012,Aiello:2015}, 
\begin{equation}
    \mathbf{E}(\rho,t)  \approx   {\rm Re}\left[\left(\hat{\varepsilon}-\frac{i\left(\varepsilon_{x}x+\varepsilon_{y}y\right)}{z_{0}}\hat{\mathbf{z}}\right)f(\mathbf{r})e^{i\left(kz-\omega t\right)}\right].  
\label{eq:E-1st-order}
\end{equation}

We note that Eq.~\eqref{eq:E-1st-order} is a good approximation to the solution to the Helmholtz equation for the regime considered in this paper. However, in an extremely tightly-focused regime one should instead use the exact electric field patterns which can be obtained numerically \cite{Novotnoy:2012,Spreeuw:2020}.

We use the resulting electric field in the atom-light Hamiltonian which drives the dipole (E1) and quadrupole (E2) transitions, 
\begin{align}
    H_{\text{AL}}(\hat{r})&=H_{E1}+H_{E2}\nonumber\\
    &=\hat{r}_i E_i + \hat{r}_i \hat{r}_j \partial_i E_j
    \label{ham}
\end{align}
where we have used Einstein's summation convention and set the electron's charge to 1. Here \(\bold{\hat{r}}\) is the position operator of the electron relative to the center-of-mass of the atom, and \(E_i\) is the electric field component of the light along the $i$-axis evaluated at the center-of-mass.

It is convenient to express \(\hat{\mathbf{r}}\) in spherical components, because its matrix elements are then easily evaluated as Clebsch-Gordan coefficients. Using the definitions, 
\begin{equation}
    \hat{r}_{\pm1}=\mp\frac{1}{\sqrt{2}}(\hat{x}\pm i\hat{y}),\qquad \hat{r}_0=\hat{z},
    \label{eq:rSph}
\end{equation}
the dipole coupling Hamiltonian can be written as, 
\begin{eqnarray*}
    H_{E1}&=& \sum_q (-1)^q \hat{r}_q E_{-q}\\
    &=& -\hat{r}_1 E_{-1}+\hat{r}_0 E_0-\hat{r}_{-1}E_1\\
    &=& \hat{r}_1\frac{-E_x+iE_y}{\sqrt{2}}+\hat{r}_0 E_z+\hat{r}_{-1}\frac{E_x+iE_y}{\sqrt{2}}, 
\end{eqnarray*}
where $\bra{J_e,m_e}\hat{r}_q\ket{J_g,m_g}\propto C_{J_g,m_g,1,q}^{J_e,m_e=m_g+q}$, and
\begin{equation}
    E_{\pm 1}=\mp\frac{1}{\sqrt{2}}(E_x\pm iE_y),\qquad E_0=E_z.
\end{equation}

Next, we evaluate the quadrupole coupling Hamiltonian. The quadrupole coupling is the product of the dyadic $r_i r_j\equiv Q_{ij}$ and the field gradient tensor $\partial_i E_j=(\nabla\mathbf{E})_{ij}$. Both may be decomposed into their irreducible spherical tensor components, $\nabla\mathbf{E}=(\nabla\mathbf{E})^{(0)}+(\nabla\mathbf{E})^{(1)}+(\nabla\mathbf{E})^{(2)}$, of rank 0, 1, and 2, respectively. A similar decomposition can be performed for the dyadic $Q$.
We note that by symmetry $Q^{(1)}=0$, and $(\nabla\mathbf{E})^{(0)}_0=-(\nicefrac{1}{\sqrt{3}})\partial_iE_i=0$ as required by the Maxwell's equations. Thus, in the E2 coupling only the rank-2 components remain and we find \cite{James:1998} \footnote{This is in disagreement with Verde {\em et al.} \cite{verde2023trapped}. Here we state that E2 coupling is only described by the rank-two tensor $Q^{(2)}$, also for $\Delta J=1$. This is in contrast with  Eq. (11) in \cite{verde2023trapped}, which we state vanishes.}, 
\begin{equation}
    H_{E2}=\sum_q(-1)^q Q^{(2)}_q (\nabla\mathbf{E})^{(2)}_{-q}. 
\label{eq:HE2a}
\end{equation}
The gradient is expressed in Cartesian components as \cite{Weissbluth:2012}, 
\begin{align}
    (\nabla\mathbf{E})^{(2)}_{\pm2}&=\frac{1}{2}(\partial_x\pm i\partial_y)(E_x\pm iE_y)\\
    (\nabla\mathbf{E})^{(2)}_{\pm1}&=\frac{1}{2}\left[\mp \partial_z(E_x\pm iE_y)\mp (\partial_x\pm i\partial_y)E_z\right]\\
    (\nabla\mathbf{E})^{(2)}_{0}&=\frac{\sqrt{6}}{2}\partial_z E_z, 
    \label{quadrupole}
\end{align}
where we have used $\partial_i E_i=0$ to simplify the equation for $(\nabla\mathbf{E})^{(2)}_{0}$. 
Finally, in the irreducible spherical tensor components representation, Eq. \eqref{eq:HE2a} can be written as
\begin{eqnarray}
H_{\mathrm{E2}} & = & \frac{1}{2}\left\{ Q_{2}^{(2)}(\partial_{x}-i\partial_{y})(E_{x}-iE_{y})\right.\nonumber\\
&  & +Q_{1}^{(2)}\left[-\partial_{z}(E_{x}-iE_{y})-(\partial_{x}-i\partial_{y})E_{z}\right]\nonumber\\
&  &+Q_{0}^{(2)}\sqrt{6}\,\partial_{z}E_{z}+\nonumber\\
 &  & +Q_{-1}^{(2)}\left[\partial_{z}(E_{x}+iE_{y})+(\partial_{x}+i\partial_{y})E_{z}\right]\nonumber\\
 &  & +\left. Q_{-2}^{(2)}(\partial_{x}+i\partial_{y})(E_{x}+iE_{y})\right\}, 
 \label{eq:quadrupole hamiltonian}
\end{eqnarray}
where $\bra{J_e,m_e}Q^{(2)}_q\ket{J_g,m_g}\propto C_{J_g,m_g,2,q}^{J_e,m_e=m_g+q}$.

We choose the magnetic field orientation $\mathbf{B}$ and the polarization, $\hat{\varepsilon}$, and the laser frequency such that $\ket{0}$ and $\ket{1}$ are resonant. Then, the quadrupole Rabi frequency (in angular frequency units) between these two states is given by 

\begin{equation}
\Omega_{E2,m_g,m_e}=\frac{e a_0^2}{\hbar}Q_{\rm red}C_{J_g,m_g,2,q}^{J_e,m_e}(-1)^q(\Tilde{\nabla}\mathbf{E})^{(2)}_{-q},
\label{eq:rabi freq}
\end{equation}
with $Q_{\text{red}}=\frac{\mel{J_e}{|Q^{(2)}|}{J_g}}{\sqrt{2J_e+1}}$ the reduced quadrupole moment in atomic units \cite{Kreuter:2005}, $e$ the elementary charge, $a_0$ the Bohr radius. We use $\Tilde{\nabla}$ to denote the gradient in the coordinates specified by the quantization axis set by $\mathbf B$. The field gradients $(\Tilde{\nabla}\mathbf{E})^{(2)}_{-q}$ define the spatial profile of the Rabi frequency. In Fig. \ref{fig1}(c) we find the relative amplitude of $(\Tilde{\nabla}\mathbf{E})^{(2)}_{-q}$ (normalized between 0 and 1) at the beam center, as a function of the angle between the magnetic field direction and propagation vector \(\vec{k}\) (\(\phi\)) and the angle between the linear polarization \(\hat{\varepsilon}\) and the magnetic field projection on the transverse plane ($\theta$) \cite{Roos:2000}. We find that the \emph{relative} coupling strength does not change in the presence of non-paraxial effects, and has no dependence on the beam waist. However, we do note a small change in the amplitude due to non-paraxial effects at the beam center.

The last term in the Hamiltonian~\eqref{hamiltonian1}includes the effect of off-resonant dipole coupling and is given by,  
 \begin{align}
     H_{\alpha}=&-\alpha_sE_0^2-\frac{\alpha_v}{J}i(\textbf{E}^{-}\times\textbf{E}^{+})\cdot\bold{\hat{J}}\nonumber\\\
     &-\frac{3\alpha_t}{J(2J-1)}\left(\frac{1}{2}\{\textbf{E}^+\cdot\bold{\hat{J}},\textbf{E}^
     -\cdot\bold{\hat{J}}\}-\frac{1}{3}J(J+1)E_0^2\right)\nonumber
 \end{align}
where \(\{,\}\) is the anticommutator and \(\bold{\hat{J}}\) are the angular momentum operators defined previously \cite{Cooper:2018}.
Here we have used the reduced dipole moments in Ref. \cite{Safranova:2011} to calculate the scalar, vector and tensor polarizability (\(\alpha_s,\alpha_v,\alpha_t\) respectively) of each $m_j$ state. 

\begin{equation}
    H_1=\tilde{H}_{\text{AL}} + H_\text{Z} + \tilde{H}_{\alpha}.
    \label{eq:full ham}
\end{equation}
where the tilde indicates $H_{\text{AL}}$ and $H_{\alpha}$ have been transformed to the quantization axis.

\subsection{Simulated light potentials}

We use second-order perturbation theory to go from $H_1$ to $H_2$, which acts on the qubit basis \(\{\ket{0},\ket{1}\}\), 
\begin{equation}
    H_2=
    \hbar\begin{pmatrix}
        \Delta/2 + \delta_0 & \Omega\\
        \Omega^* & -\Delta/2+\delta_1
    \end{pmatrix}
    \label{eq:2x2ham}
\end{equation}
Here, the diagonal energy shifts are the AC stark shifts on the qubit states due to the dipole and quadrupole coupling to other $m_j$ states, given by \(\delta_i=\sum_k\Omega_{ik}\Omega_{ki}/\Delta_{ik}\), where \(\Delta_{ik}=E_i-E_k\) is the energy difference between the qubit states $i=0,1$ and other $m_j$ states $k=2\ldots 7$. Furthermore, off-resonant Raman coupling slightly shifts the Rabi frequency between the qubit states by $\Omega_{s}=\sum_k\Omega_{0k}\Omega_{k1}/\Delta_{kl}$ where $l$ labels the qubit state in the same manifold as state $k$. We include this effect in Eq.~\eqref{eq:2x2ham} via the definition of the Rabi frequency between the qubit states $\Omega=\Omega_{01}+\Omega_s$. We note that the two-level approximation to $H_1$ remains valid for the range of laser parameters considered here.

The eigenvalues of $H_2$ give the light potentials $E(x)$ in the dressed basis $\{\ket{+},\ket{-}\}$, where $\ket{+}=c_1\ket{0}+c_2\ket{1}$ and $\ket{-}=c_1\ket{1}-c_2\ket{0}$. For $B_x$ the potential is suppressed and spatially-symmetric about the origin. For $B_y$ we note that while the profile is Gaussian, it is shifted from the atom's center-of-mass position. Moreover, we plot the potentials for a Laguerre-Gauss mode \(\text{LG}_{p,l}=\text{LG}_{01}\) in Fig. \ref{potential plots}(e)-(h), and observe a similar shifted potential in the $B_y$ case. 

\begin{figure}[t]
    \centering
    \includegraphics[width=0.9\linewidth]{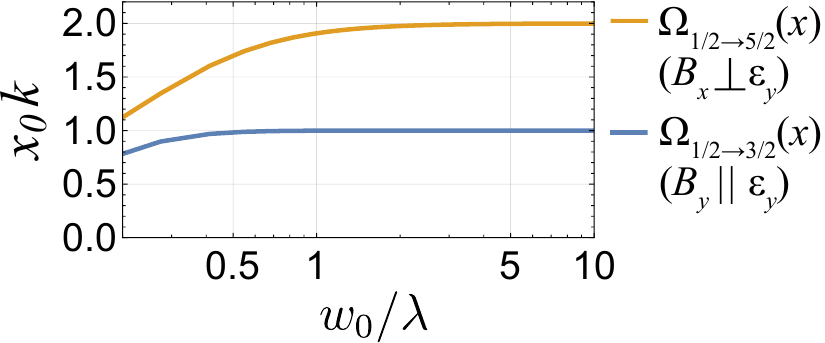}
    \caption{\textbf{Rabi frequency displacement.} Displacement $x_0$ of the peak of the Rabi frequency $\Omega(x)$ as a function of waist size $w_0$. The scale is logarithmic on the horizontal axis. When $\bold{B}\parallel\varepsilon_y$, the resonant coupling $\Omega_{1/2\rightarrow3/2}(x)$ is displaced by $x_0=\lambda/2\pi$ (blue line) in the limit $w_0\gg\lambda$. When $\bold{B}\perp\varepsilon_y$ the resonant coupling $\Omega_{1/2\rightarrow5/2}(x)$ is displaced by $x_0=\lambda/\pi$ (orange line) in the same limit.}
    \label{magnus effect displacements}
\end{figure}

In Fig.~ \ref{magnus effect displacements} we determine the position of the peak ($x_0$) of $\Omega(x)$ to characterize how $x_0$ shifts in the direction of the focal plane. We consider the case of (i) $\mathbf{B}= B_y \parallel \varepsilon_y$ and (ii) $\mathbf{B}=B_x\perp \varepsilon_y$. In the first case, the coupling $\Omega_{E2, 1/2, 3/2} (x)$ is resonant and $x_0$ approaches $k^{-1}=\lambda/2\pi$ as $w_0\gg\lambda$, as shown in Fig. \ref{magnus effect displacements}. In the second case, the displacement of resonant coupling $\Omega_{E2, 1/2, 5/2} (x)$  approaches $2k^{-1}=\lambda/\pi$ as $w_0\gg\lambda$ (see Fig. \ref{magnus effect displacements}). In both cases, for very small beam waists $w_0\ll\lambda$ (which are typically below the diffraction limit) $x_0$ decreases. Finally, we note when $q$ is negative the peak is displaced to the left instead of to the right (i.e. for $\Omega_{E2,1/2,-1/2}$, $x_0$ approaches $-\lambda/2\pi$ as $w_0\gg\lambda$). 

We plot the the spatial profile of the AC Stark shifts \(\delta_i\) for $\mathbf B= B_y \hat y$ in \ref{potential plots}. The Stark shifts for $\mathbf B=B_x \hat x$ are negligible in comparison. We note that the spatial profile is different from the Rabi frequency and in particular, the displacement is in the opposite direction. This is due to the dominant contribution to $\delta_i$ ($\Omega_{E2,1/2,-1/2}$) being centered at $x\approx-\lambda/2\pi$. The Stark shift scales with the square of the Rabi frequency amplitude, $\delta\propto|\Omega|^2$, and is inversely proportional to the magnetic field strength \(|\bold{B}|\). For an atom positioned at the peak of the Rabi frequency, this displacement results in a Stark shift gradient at the atom's position.

\subsection{Single-qubit infidelity}

The largest contribution to the infidelity in Eq. \eqref{analytic fid} is $\kappa_x$, which is predominantly due to the Rabi frequency curvature \(\Omega^{(2)}\) when \(\Omega^{(2)}/\Omega^{(0)}>2(\delta^{(1)})^2/\omega^2\). We note that the ratio of the Rabi frequency curvature to its amplitude $\Omega^{(2)}/\Omega^{(0)}$ changes geometrically, for instance as the beam waist is decreased. This in turn increases the Stark shifts quadratically which means that for sufficiently small trap frequency \(\omega\) or magnetic field strength, and at large enough beam intensities, non-paraxial effects are the dominant source of error. One such choice of parameters is \(\Omega^{(0)}=2\pi~1\)~MHz, $\bold{B}=2.5$~G and $\omega=2\pi~40$~kHz, where the non-paraxial $\delta^{(1)}$ error surpasses the $\Omega^{(2)}$ error. However one should keep in mind that these parameters are unusual for trapped ion experiments. 

We partially compensate the $\kappa_x$ error by using a gate time $\Bar{t}_g=\pi/2\Bar{\Omega}$, where 
\begin{align}
\Bar{\Omega}=&\Omega^{(0)}+\Omega^{(2)}l_{\text{ho}}^2(2\Bar{n}+1)\notag\\
&+\left(2\Omega^{(0)}(\delta^{(1)})^2 l_{\text{ho}}^2/\omega^2\right)(2\bar{n}+1)
\end{align}
is the thermally-averaged Rabi frequency. The first correction, $\Omega^{(2)}l_{\text{ho}}^2(2\Bar{n}+1)$, is due to the Rabi frequency curvature. The second correction, $2\Omega^{(0)}(\delta^{(1)})^2l_{\text{ho}}^2/\omega^2$, is due to the Stark shift gradient, only occurring in the non-paraxial case. Using the new gate time, we expand Eq.~\eqref{fidelity} to second order in \(\kappa_{x/z}\) to arrive at the infidelity of the compensated gate,  
\begin{align}
    1-&\Bar{F}_{\Bar{t}}\approx \frac{2}{3}\cos^2(\frac{\pi\gamma}{2})-(1+2\Bar{n})\frac{\pi\kappa_x}{3\Bar{\Omega}}\sin(\pi\gamma)\notag\\
    &-\frac{1}{6(\Omega^{(0)})^2}(1+8\Bar{n}(\Bar{n}+1))\biggl(\pi^2\kappa_x^2\gamma^2\cos(\pi\gamma)\notag\\
    &\hskip30pt +\frac{\kappa_z^2}{2}\left(\cos(\pi\gamma)-1\right)+\frac{\pi\kappa_z^2\gamma}{4}\sin(\pi\gamma) \biggr)
    \label{analytic fid appendix 2}
\end{align}
where \(\gamma=\Omega^{(0)}/\Bar{\Omega}\). 

\subsection{Parallel single-qubit gates}

We consider addressing two ions in an $N$-ion chain simultaneously in order to perform parallel single-qubit gates. In the interaction picture with respect to $H_0=\omega_m \hat{a}^{\dagger}_m\hat{a}_m$, the Hamiltonian is:
\begin{align}&H_4=\Omega^{(0)}(\hat{\sigma}_x^{(i)}+\hat{\sigma}_x^{(j)})+\Omega^{(2)}\left(\hat{x}_i^2\hat{\sigma}_x^{(i)}+\hat{x}_j^2\hat{\sigma}_x^{(j)}\right)\nonumber\\
&+\delta^{(1)}\left(\hat{x}_i\hat{\sigma}_z^{(i)}+\hat{x}_j\hat{\sigma}_z^{(j)}\right)+\delta^{(2)}\left(\hat{x}_i^2\hat{\sigma}_z^{(i)}+\hat{x}_j^2\hat{\sigma}_z^{(j)}\right).
\label{eq:2qubitham}
\end{align}
The first term of Hamiltonian \eqref{eq:2qubitham} describes parallel $\hat{\sigma}_x$ qubit rotations. Performing a Lang-Firsov transformation on $H_4$ with unitary $\hat{U}_1=\exp(\sum_m\zeta_m(b_{mi}\hat{\sigma}_z^{(i)}+b_{mj}\hat{\sigma}_z^{(j)})(\hat{a}^{\dagger}_m-\hat{a}_m))$ where $\zeta_m=\delta^{(1)}l_m/\omega_m$ eliminates the linear phonon coupling. Disregarding global energy offsets and off-resonant processes that change the Fock state, the transformed Hamiltonian is
\begin{align}
    H_5=&\Omega^{(0)}(1+2\zeta_m^2(2\hat{n}_m+1))(b_{mi}\hat{\sigma}_x^{(i)}+b_{mj}\hat{\sigma}_x^{(j)})\nonumber\\
    &+\Omega^{(2)}l_m^2(2\hat{n}_m+1)(b_{mi}\hat{\sigma}_x^{(i)}+b_{mj}\hat{\sigma}_x^{(j)})\nonumber\\
    &+\delta^{(2)}l_m^2(2\hat{n}_m+1)(b_{mi}\hat{\sigma}_z^{(i)}+b_{mj}\hat{\sigma}_z^{(j)})\nonumber\\
    &-2\omega_m\zeta_m^2b_{bi}b_{mj}\hat{\sigma}_z^{(i)}\hat{\sigma}_z^{(j)}.
    \label{eq:ham4}
\end{align} 
Here we have used the Baker-Haussdorf lemma to determine the leading-order corrections as in the single qubit case (see the Supplemental Material). This Hamiltonian describes qubit-motion coupling of the same form as the single-qubit case, albeit now coupling to each axial mode. We note an additional qubit-qubit interaction (final term), making operations on qubit $i$ dependent on the state of qubit $j$.

\section*{Acknowledgements}
We thank Liam J. Bond for help on the Hamiltonian transformations. This work was supported by the Netherlands Organization for Scientific Research (Grant Nos. 680.91.120, VI.C.202.051 and 680.92.18.05), the Dutch Research Council (Grant No. OCENW.M.22.403), and by the Horizon Europe programme HORIZON-CL4-2021-DIGITAL-EMERGING-01-30 via project 101070144 (EuRyQa) A.S.N. is supported by the Dutch Research Council (NWO/OCW) as part of the Quantum Software Consortium programme (project number 024.003.037). A.S.N. is supported by Quantum Delta NL (project number NGF.1582.22.030).

\section{Author contributions}
LG performed the analytic calculations and wrote the manuscript with support from RS. LG and MM performed the numerical simulation of gate errors. ZA supported the numerical simulations and interpretation of results. RG, ASN and RS conceptually developed and supervised the project. All authors contributed to the manuscript.

\bibliography{paper_bib}

\clearpage

\renewcommand{\theequation}{S\arabic{equation}}
\renewcommand{\thefigure}{S\arabic{figure}}
\renewcommand{\thesubsection}{S\arabic{subsection}}
\renewcommand{\thetable}{S\arabic{table}}

\section{Supplemental Material}

\subsection{Gradient and curvature terms}
\label{appendix_terms}
Here we give analytic expressions for the gradient and curvature terms in $H_3$ which cause the considered errors in the main text. The Rabi frequency $\Omega(x)$ is approximately a displaced Gaussian, and we make a Taylor expansion about the Rabi frequency peak $x_0$ (where the ion is positioned):
\begin{align}
\Omega(\hat{x})&\approx\Omega^{(0)}\exp\left(-\frac{(\hat{x}-x_0)^2}{w_0^2}\right)\nonumber\\
    &\approx\Omega^{(0)}\left(1-\frac{(\hat{x}-x_0)^2}{w_0^2}+\dots \right).
\end{align}
We identify the coefficient of $\hat{x}^2$ as: $\Omega^{(2)}\approx\frac{\Omega^{(0)}}{w_0^2}$. Similarly, the Stark shift is a displaced Gaussian but now shifted in the opposite direction and centered at $-x_0$. The Taylor expansion about the ion's position yields:
\begin{align}
\delta(\hat{x})&\approx\delta^{(0)}\exp\left(-\frac{(\hat{x}+x_0)^2}{w_0^2}\right)\nonumber\\
    &\approx\delta^{(0)}\exp(-4x_0^2/w_0^2)\biggl(1-\frac{4x_0}{w_0^2}(\hat{x}-x_0)~\nonumber\\
    &~~-\frac{(w_0^2-8x_0^2)}{w_0^4}(\hat{x}-x_0)^2+\dots \biggr).
\end{align}
For larger waists with $w_0^2\gg x_0^2$ such that $\exp(-4x_0^2/w_0^2)\sim1$, the gradient and curvature terms simplify to $\delta^{(1)}\sim\frac{4\delta^{(0)}x_0}{w_0^2}$ and $\delta^{(2)}\sim\frac{\delta^{(0)}(w_0^2-8x_0^2)}{w_0^4}$. Note that in our case with $w_0=\lambda$ and $x_0=\lambda/2\pi$,  $\exp(-4x_0^2/w_0^2)=\exp(-\pi^{-2})\sim0.9$.

\subsection{Hamiltonian transformations}
\label{transformations}

We perform a qubit-dependent squeezing transformation on $H_3$ with unitary 
\begin{align}
    \hat{S}(s\hat{\sigma}_z) = \exp(\frac{s\hat{\sigma}_z}{2}(\hat{a}^2 - \hat{a}^{\dagger}{}^2)),
\end{align}
Using the quadrature transformations $\hat{S}^\dagger(s\hat{\sigma}_z) \hat{a} \hat{S}(s\hat{\sigma}_z) = \hat{a} \cosh(s\hat{\sigma}_z) - \hat{a}^\dagger \sinh(s\hat{\sigma}_z)$ and $\hat{S}^\dagger(s\hat{\sigma}_z) \hat{a}^\dagger \hat{S}(s\hat{\sigma}_z) = \hat{a}^\dagger \cosh(s\hat{\sigma}_z) - \hat{a} \sinh(s\hat{\sigma}_z)$, 
the motional terms transform as
\begin{subequations}
\begin{align}
    \hat{S}^\dagger(s\hat{\sigma}_z) \hat{x}^2 \hat{S}s\hat{\sigma}_z) &= e^{-2s\hat{\sigma}_z} \hat{x}^2, \\ 
    \hat{S}^\dagger(s\hat{\sigma}_z) \hat{x} \hat{S}(s\hat{\sigma}_z) &= e^{-s\hat{\sigma}_z} \hat{x}, \\ 
    \hat{S}^\dagger(s\hat{\sigma}_z) \hat{p}^2 \hat{S}(s\hat{\sigma}_z) &= e^{2s\hat{\sigma}_z} \hat{p}^2. 
\end{align}
\end{subequations}
The squeezed Hamiltonian is
\begin{align}
    \tilde{H} &= \hat{S}^\dagger(s\hat{\sigma}_z) H_3 \hat{S}(s\hat{\sigma}_z) \nonumber\\
    &= \frac{1}{2m} e^{2s\hat{\sigma}_z} \hat{p}^2 + \frac{1}{2}e^{-2s\hat{\sigma}_z} m \omega^2 \hat{x}^2\nonumber\\
    &+ \hbar\left( \delta^{(0)} +\delta^{(1)} e^{-s\hat{\sigma}_z}\hat{x} + \delta^{(2)} e^{-2s\hat{\sigma}_z} \hat{x}^2 \right) \hat{\sigma}_z \nonumber\\
    &+ \hbar \hat{S}^\dagger(s\hat{\sigma}_z) \hat{\Omega}\hat{\sigma}_x \hat{S}(s\hat{\sigma}_z)
\end{align}
We define a qubit-dependent frequency $\Tilde{\omega} = \sqrt{\omega^2 + \frac{2\delta^{(2)} \hbar}{m}\hat{\sigma}_z}$, which absorbs $\delta^{(2)}$ into the ion's harmonic motion. By 1st-order Taylor expansion $\tilde{\omega}\approx\omega+(\delta^{(2)}l_{\text{ho}}^2)\hat{\sigma}_z$. The Rabi frequency components are denoted \(\hat{\Omega}=\Omega^{(0)} + \Omega^{(2)} \hat{x}^2\). In terms of creation and annihilation operators, the Hamiltonian (in angular frequency units) is
\begin{align}
    \tilde{H} = &-\frac{e^{2s\hat{\sigma}_z}}{4}\omega \left(\hat{a}^{\dagger^2}+\hat{a}^2-2\hat{a}^{\dagger}\hat{a}-1\right)\nonumber \\
    &+ \frac{e^{-2s\hat{\sigma}_z}}{4} \frac{\tilde{\omega}^2}{\omega} \left(\hat{a}^{\dagger^2}+\hat{a}^2+2\hat{a}^{\dagger}\hat{a}+1\right)\notag\nonumber\\
    &+ \left( \delta^{(0)} +\delta^{(1)} e^{-s\hat{\sigma}_z}l_{\text{ho}}(\hat{a}+\hat{a}^{\dagger}) \right) \hat{\sigma}_z\nonumber\\
    &+  \hat{S}^\dagger(s\hat{\sigma}_z) \hat{\Omega}\hat{\sigma}_x \hat{S}(s\hat{\sigma}_z)
\end{align}

We recover the harmonic oscillator with frequency $\tilde{\omega}$ by choosing the squeezing parameter \(s =\delta^{(2)}l_{\text{ho}}^2/\omega\) as a dimensionless measure of the Stark shift curvature. The Hamiltonian is then
\begin{align}
    \tilde{H} = &\Tilde{\omega} (\hat{a}^\dagger \hat{a}+1/2) + \delta^{(0)}\hat{\sigma}_z\nonumber \\
    &+ \delta^{(1)} e^{-{s\hat{\sigma}_z}}x_{0}(\hat{a}+\hat{a}^{\dagger})\hat{\sigma}_z
    \nonumber\\
    &+ \hat{S}^\dagger(s\hat{\sigma}_z)\hat{\Omega}\hat{\sigma}_x \hat{S}(s\hat{\sigma}_z).
\end{align}
Similarly the linear motion term can be transformed through a qubit-dependent Lang-Firsov transformation with unitary 
\begin{align}
\hat{D}(\zeta\hat{\sigma}_z)=\exp(\zeta\hat{\sigma}_z(\hat{a}^{\dagger}-\hat{a})),
\end{align}
with quadrature transformations $\hat{D}^\dagger(\zeta\hat{\sigma}_z) \hat{a} \hat{D}(\zeta\hat{\sigma}_z) = \hat{a} - \zeta\hat{\sigma}_z$ and $\hat{D}^\dagger(\zeta\hat{\sigma}_z) \hat{a}^{\dagger} \hat{D}(\zeta\hat{\sigma}_z) = \hat{a}^{\dagger} - \zeta^*\hat{\sigma}_z$. The Hamiltonian becomes
\begin{align}
    \tilde{H}=&\tilde{\omega}\left(\hat{a}^\dagger \hat{a}+1/2-\zeta\hat{\sigma}_z(\hat{a}+\hat{a}^{\dagger})+\zeta^2\right)\nonumber\\
    &+\delta^{(0)}\hat{\sigma}_z+\delta^{(1)} e^{-{s\hat{\sigma}_z}}x_{0}(\hat{a}+\hat{a}^{\dagger})\hat{\sigma}_z\nonumber\\ 
    &-2\delta^{(1)} e^{-{s\hat{\sigma}_z}}x_{0}\zeta\nonumber\\
    &+  \hat{D}^{\dagger}(\zeta\hat{\sigma}_z)\hat{S}^\dagger(s\hat{\sigma}_z)\hat{\Omega} \hat{\sigma}_x \hat{S}(s\hat{\sigma}_z)\hat{D}(\zeta\hat{\sigma}_z),
\end{align}
when $\zeta$ is real. The linear motion terms cancel out by choosing the displacement parameter $\zeta=\delta^{(1)}l_{\text{ho}}/\omega$ to be a dimensionless measure of the Stark shift gradient. The Hamiltonian simplifies to 
\begin{align}
\tilde{H}=&\Tilde{\omega} (\hat{a}^\dagger \hat{a}+1/2) - \Tilde{\omega}\zeta^2+ \delta^{(0)}\hat{\sigma}_z \nonumber
\\
&+  \hat{D}^{\dagger}(\zeta\hat{\sigma}_z)\hat{S}^\dagger(s\hat{\sigma}_z)\hat{\Omega} \hat{\sigma}_x \hat{S}(s\hat{\sigma}_z)\hat{D}(\zeta\hat{\sigma}_z).
\end{align}
Disregarding energy offsets that do not contribute to qubit-motion coupling, like $\delta^{(0)}\hat{\sigma}_z$ which can be fixed by the laser frequency, the Hamiltonian is
\begin{align}
    \tilde{H}=&\tilde{\omega}(\hat{a}^{\dagger}\hat{a}+1/2)\nonumber\\
    &+ \hat{D}^{\dagger}(\zeta\hat{\sigma}_z)\hat{S}^\dagger(s\hat{\sigma}_z) \hat{\Omega}\hat{\sigma}_x \hat{S}(s\hat{\sigma}_z)\hat{D}(\zeta\hat{\sigma}_z).
    \label{analytic ham}
\end{align}
The first term side describes the ion's motion as a modified harmonic oscillator with frequency \(\tilde{\omega}\).
The second term describes the ion-laser coupling acted on by squeeze and displacement operators dependent on the qubit state of the ion. We evaluate the second term using the Baker-Hausdorff (BH) lemma, which gives corrections to the Rabi frequency amplitude $\Omega^{(0)}$. Disregarding off-resonant processes that change the Fock state,
\begin{align}
    \hat{D}^{\dagger}(\zeta&\hat{\sigma}_z)\hat{S}^\dagger(s\hat{\sigma}_z) \left(\Omega^{(0)}\hat{\sigma}_x\right) \hat{S}(s\hat{\sigma}_z)\hat{D}(\zeta\hat{\sigma}_z)\label{eq:rabi_amplitude}\\
    &=\Omega^{(0)}\left(1+2\zeta^2(2\hat{n}+1)+\frac{s\zeta}{2}(2\hat{n}+1)+\order{\hat{n}^2}\right)\hat{\sigma}_x.
    \nonumber
\end{align}
The second term proportional to $(s\zeta/2)$ is suppressed by $10^2$ and thus negligible, along with higher order terms.

We then calculate the squeeze and displacement transformation on the Rabi frequency curvature $\Omega^{(2)}\hat{x}^2$, which is expressed in creation and annihilation operators as $\Omega^{(2)}\hat{x}^2\hat{\sigma}_x=\Omega^{(2)}l_{\text{ho}}^2(2\hat{n}+1+\hat{a}^{\dagger}\hat{a}^{\dagger}+\hat{a}\hat{a})\hat{\sigma}_x$. Again using the BH lemma, we find
\begin{widetext}
\begin{align}
    \hat{D}^{\dagger}(\zeta\hat{\sigma}_z)\hat{S}^\dagger(s\hat{\sigma}_z) \left(\Omega^{(2)}\hat{x}^2\hat{\sigma}_x\right) \hat{S}(s\hat{\sigma}_z)\hat{D}(\zeta\hat{\sigma}_z)&=\Omega^{(2)}l_{\text{ho}}^2(2\hat{n}+1)\left(1+2\zeta^2(2\hat{n}+1)+\frac{s\zeta}{2}(2\hat{n}+1)+\order{\hat{n}^2}\right)\hat{\sigma}_x
    \label{eq:rabi_curvature}
\end{align}
The second term proportional to $\zeta^2$ is suppressed by $\sim10^6$, so we disregard it along with higher order terms. Inserting Eq. \eqref{eq:rabi_amplitude} and Eq. \eqref{eq:rabi_curvature}, the Hamiltonian is written in terms of the physical parameters as
\begin{align}
    \tilde{H}=&\omega (\hat{a}^{\dagger}\hat{a}+1/2) + \Omega^{(0)}\hat{\sigma}_x+
    2\delta^{(2)}l_{\text{ho}}^2 (\hat{n}+1/2) \hat{\sigma}_z
    +\left( (2\Omega^{(0)}(\delta^{(1)})^2l_{\text{ho}}^2/\omega^2)(2\hat{n}+1)+\Omega^{(2)}l_{\text{ho}}^2(2\hat{n}+1) \right)\hat{\sigma}_x.
    \label{ham simple}
\end{align}

\end{widetext}
In the text this is written in the following parts 
\begin{equation}
\tilde{H}=H_{\text{id}}+\kappa_z(\hat{n}+1/2)\hat{\sigma}_z+\kappa_x(2\hat{n}+1)\hat{\sigma}_x
\end{equation}
where 
\begin{align}
    H_{\text{id}}&=\omega(\hat{a}^{\dagger}a+1/2)+\Omega^{(0)}\hat{\sigma}_x,\\
    \kappa_z&=2\delta^{(2)}l_{\text{ho}}^2,\\
    \kappa_x&=2\Omega^{(0)}(\delta^{(1)})^2l_{\text{ho}}^2/\omega^2+\Omega^{(2)}l_{\text{ho}}^2.
\end{align}

In our analytic model we have neglected off-resonant processes which change the Fock state $n$. In Fig. 3 we see that off-resonant terms become significant when $\bar{n}\ll 1$ or $\Omega\gg\omega$, by comparing the analytic and numerical curves. We note that instead of using the BH lemma we can utilize the expression for the overlap of a squeezed-displaced Fock state from Ref. \cite{mo1996displaced}. The overlap gives the transformed Rabi frequency for a given Fock state, \(\Omega^{(0)}\braket{-\zeta,-s,n}{\zeta,s,n}\hat{\sigma}_x\), which is well approximated by the first few terms of the BH lemma when $\zeta$ and $s$ are small. The overlap also allows the calculation of off-resonant processes that change the Fock state $n\rightarrow m$, by \(\Omega^{(0)}\braket{-\zeta,-s,m}{\zeta,s,n}\hat{\sigma}_x\). This could be of interest, for instance to find tweezer parameters that specifically drive the sidebands $n\rightarrow n\pm1$ or generate anharmonic potentials.

\subsection{Dependence on tweezer waist}
\label{appendix_waist}

\begin{figure}[h]
    \centering
    \includegraphics[width=\linewidth]{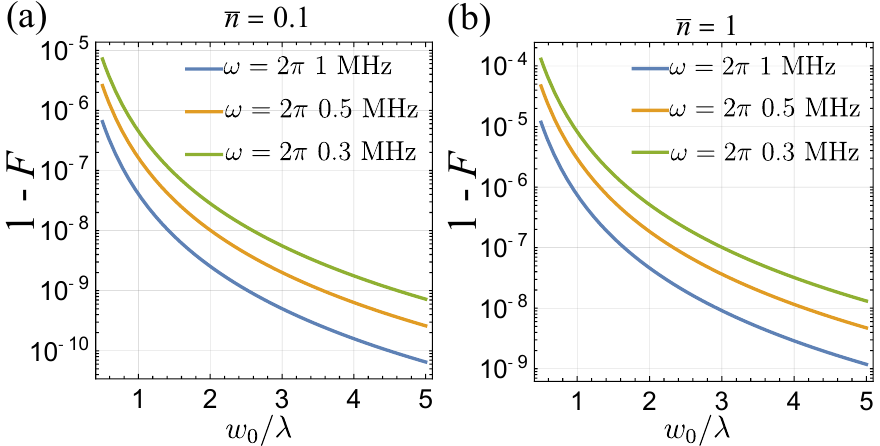}
    \caption{Analytic infidelity from Eq. \ref{analytic fid appendix 2} for the single qubit gate as a function of tweezer waist $w_0$, with different trap frequencies $\omega$. Here $\Omega^{(0)}=2\pi~0.2$~MHz and $\bold{B}=2.5$~G, and the ion temperature is fixed at (a) $\bar{n}=0.1$ and (b) $\bar{n}=1$.}
    \label{fig:fid_vs_waist}
\end{figure}

We consider how the single qubit infidelity (Eq. (20)) varies with the beam waist $w_0$ in Fig. \ref{fig:fid_vs_waist}. As the waist increases non-paraxial errors become practically negligible. However, we note that increasing the waist is not always a valid solution in experimental setups - for instance when addressing multiple ions in a chain. Here, a larger waist can cause cross talk errors with adjacent ions \cite{Parrado:2021}. In this case, one could find an optimal tweezer waist which minimizes both cross talk and non-paraxial effects. A more straightforward solution would be to use a magnetic field and polarization orientation that suppresses the non-paraxial error.

\end{document}